%% file: samplepaper.tex
\documentclass[runningheads]{llncs}

\let\llncssubparagraph\subparagraph
\let\subparagraph\paragraph
\usepackage[compact]{titlesec}
\let\subparagraph\llncssubparagraph

\usepackage{graphicx}
\usepackage{enumitem}
\setlist{nolistsep}

\usepackage{float}

\usepackage{booktabs}
\usepackage{CJKutf8}

\newcommand{\chinese}[1]{\begin{CJK}{UTF8}{bsmi}#1\end{CJK}}

\usepackage[numbers,sort&compress]{natbib}
\newcommand{\pro}{\texttt{pro}\space}
\newcommand{\anti}{\texttt{anti}\space}
\newcommand{\unsure}{\texttt{unsure}\space}
\newcommand{\rom}[1]{\uppercase\expandafter{\romannumeral #1\relax}}

\usepackage{hyperref}
\usepackage{xcolor}
\hypersetup{
  colorlinks   = true, %Colours links instead of ugly boxes
  urlcolor     = blue, %Colour for external hyperlinks
  linkcolor    = black, %Colour of internal links
  citecolor   = blue %Colour of citations
}

\usepackage{color}

\begin{document}
\title{The Dynamics of Political Narratives During the Russian Invasion of Ukraine}
% \thanks{Supported by organization x.}}
%
\titlerunning{Political Narratives During Russia-Ukraine War}
% If the paper title is too long for the running head, you can set
% an abbreviated paper title here
%
\author{Ahana Biswas\inst{1}\orcidID{0000-0002-0884-8984} \and
Tim Niven\inst{2}\orcidID{0000-0003-2423-6037} \and
Yu-Ru Lin\inst{1}\orcidID{0000-0002-8497-3015}}
% \author{Anonymous Author(s)}
%
\authorrunning{A. Biswas et al.}

% First names are abbreviated in the running head.
% If there are more than two authors, 'et al.' is used.
%

\institute{University of Pittsburgh, Pittsburgh PA 15260, USA\\ 
\email{\{ahana.biswas, yurulin\}@pitt.edu}\\
\and
Doublethink Lab, Taiwan\\
% \url{http://www.springer.com/gp/computer-science/lncs} \and
% ABC Institute, Rupert-Karls-University Heidelberg, Heidelberg, Germany\\
\email{tim@doublethinklab.org}}

% \institute{}
% %

\maketitle              % typeset the header of the contribution
\vspace{-2em}
\begin{abstract}
The Russian invasion of Ukraine has elicited a diverse array of responses from nations around the globe. During a global conflict, polarized narratives are spread on social media to sway public opinion. We examine the dynamics of the political narratives surrounding the Russia-Ukraine war during the first two months of the Russian invasion of Ukraine (RU) using the Chinese Twitter space as a case study. Since the beginning of the RU, pro-Chinese-state and anti-Chinese-state users have spread divisive opinions, rumors, and conspiracy theories. We investigate how the pro- and anti-state camps contributed to the evolution of RU-related narratives, as well as how a few influential accounts drove the narrative evolution. We identify pro-state and anti-state actors on Twitter using network analysis and text-based classifiers, and we leverage text analysis, along with the users' social interactions (e.g., retweeting), to extract narrative coordination and evolution. We find evidence that both pro-state and anti-state camps spread propaganda narratives about RU. Our analysis illuminates how actors coordinate to advance particular viewpoints or act against one another in the context of global conflict.

\keywords{Computational propaganda  \and Russia-Ukraine conflict \and Social network analysis.}
\end{abstract}
\vspace{-1.5em}
\section{Introduction}
Social media is often used as a means to manipulate public opinion during an international conflict, where polarized narratives are disseminated by different groups. Since the beginning of the Russian invasion of Ukraine, pro-Chinese-state (pro-state) and anti-Chinese-state (anti-state) accounts have spread divisive opinions, rumors, and conspiracy theories~\cite{wong2022us,medium,pol}. One of the propaganda narratives employed by the pro-state actors is the Ukraine biolabs conspiracy which states that the US has set up biolabs in Ukraine to manufacture bioweapons~\cite{10020223,wong2022us,pol}. Such narratives were found to have originated from Russian social media accounts as a means to justify the invasion of Ukraine~\cite{pol}.  
Existing research \cite{10020223} found that the largest group of coordinated accounts tweeting Ukraine biolabs narratives also extensively retweeted Russian and Chinese state-funded media, suggesting certain coordinated efforts may exist between the Russian and Chinese users to sway public opinion on social media.

This study uses Chinese Twitter as a case study to examine political narratives surrounding the Russia-Ukraine war in the first two months of the Russian invasion of Ukraine (RU). Using text-based and network-driven approaches, we leverage a Twitter dataset to identify pro-state and anti-state actors. We investigate the subtopics introduced in the pro-state and anti-state networks, as well as how attention shifted toward the subtopics in subsequent weeks. We also look at how the pro- and anti-state camps contribute to the various narratives over time. We identify pro-state and anti-state influential actors based on their position in the retweet network and investigate how they influence the evolution of RU-related narratives. 
Our research questions (RQs) are:
% \\
\vspace{1.5mm}

\begin{itemize}
\item[{\bf RQ1}] What were the dominant narratives in pro- and anti-state networks during the first two months of RU?
\item[{\bf RQ2}] How did the pro- and anti-state camps contribute to the evolution of the RU-related narratives? How did influential accounts shape the evolution of the RU-related narratives?
\end{itemize}

\subsubsection{Main Contributions.} The main contributions of this work are as follows:
\begin{itemize}
    \item[(1)] By combining text analysis and network-based approaches, we propose a method for identifying pro-state vs. anti-state leaning accounts, narratives, and influential actors in large social networks.
    
    \item[(2)] This is the first large-scale analysis of narratives and interactions between pro- and anti-state camps on Chinese Twitter during the Russia-Ukraine war. We discover propagandist narratives from both the camps. Our study also reveals that anti-state voices dominate the Chinese Twitter space, which is consistent with previous studies ~\cite{bolsover2019chinese}.  
\end{itemize}

\section{Background and Related Work}
\subsubsection{Chinese Diplomatic Communication and Propaganda.}
Several works have studied the Chinese government's information censorship policies, online public opinion manipulation, and diplomatic communication strategies in digital/online space~\cite{bolsover2019chinese, huang2019building, schliebs2021china, wu2019talking}.~\citet{king_pan_roberts_2017} found that the Chinese government posts about 448M social media comments a year which are fabricated by government employees as part of their regular jobs. 
%~\cite{king_pan_roberts_2017} also show that the Chinese state's strategy is to avoid arguments with their critics. 
Media reports have suggested that the Chinese official Twitter accounts attempt to influence public opinion~\cite{collins2017twitter,kaiman2014free,phillips2014china}.~\citet{bolsover2019chinese} found that the anti-Chinese state groups predominantly manipulate information in the Chinese Twitter space. These observations regarding the distinct narratives promoted by pro-state and anti-state accounts inspired our study, which identifies the political orientations of the accounts from which the narratives originated.

\subsubsection{Analysis of Propaganda.}
Social media manipulation to shape public opinion has become a matter of growing concern~\cite{bradshaw2021industrialized}. Existing works have explored the role of coordinated behavior~\cite{pacheco2020unveiling, chetan2019corerank} and bots/trolls to drive information campaigns~\cite{ferrara2020types,ferrara2016rise, zhang2016rise}.~\citet{gonzalez2021bots} studied the visibility of bots, human accounts, and verified accounts and found that bots are less central in information campaigns than verified accounts during contentious political events. Research on propaganda detection from a text analysis perspective has focused on using distant supervision~\cite{mintz2009distant}, and detecting the use of propaganda techniques~\cite{da2019fine,da2019findings}. It has been seen that `malicious' actors change their behaviors to prevent detection by moderation tools~\cite{cresci2017paradigm}. Following the suggestion of~\citet{martino2020survey}, we leverage text-based and network analysis approaches to identify actors potentially involved in information campaigns.  

\subsubsection{Analysis of Evolving Networks, Communities and Topics.}
Previous works have explored content co-sharing networks to discover topic-based user communities and the engaging topics in these communities~\cite{ferreira2021dynamics,nobre2022hierarchical,gomes2020unveiling}.~\citet{arazzi2023importance} studied the importance of adopted language in the formation of online communities around specific topics. In this work, we use $n$-grams co-occurrence in tweets to extract fine-grained narratives and analyze the evolution of narratives in the pro-state and anti-state camps.

\section{Data}

\subsubsection{Twitter dataset.} 
We collect all tweets containing specific keywords related to Ukraine biolabs conspiracy (see \href{https://osf.io/xrb76/?view_only=a8439d2b7b8548d19b65dd4e56478aab}{Supplementary Materials}\footnote{https://osf.io/xrb76/?view\_only=a8439d2b7b8548d19b65dd4e56478aab})
from March 2022 until April 2022, i.e., during the first two months (8 weeks) of RU. This returns Ukraine biolabs-related tweets from 4494 Twitter accounts. We collect all the tweets for these 4494 accounts in the study period, resulting in a dataset containing over 4M tweets (including around 1.6M replies and 223K unique retweets).

\subsubsection{Retweet network.} 
 
We construct a retweet network using the retweets in our Twitter dataset, where nodes are unique Twitter user accounts and a directed edge $A \rightarrow B$ indicates user $A$ retweeted $B$. The edges are weighted by the number of times a source user retweeted a target user. Our retweet network comprises over 74K accounts and around 2M connections between them. We expand our data by extracting an additional 5000 accounts from the retweet network based on their impact on generating retweets, taking their immediate connections and network position into account. More specifically, we calculate an account's {\it influence} by the harmonic mean of its in-degree centrality and betweenness centrality in the retweet network.

\section{Methodology}

\subsubsection{Tweet-level classification.}
We manually annotate around 12.6K tweets by the textual content as \pro (pro-state stance), \anti (anti-state stance), or \unsure (neither pro-state nor anti-state). The inter-annotator agreement (Cohen's Kappa) is 0.7 (substantial agreement). We select a set of over 11K tweets containing specific topic-related keywords (see \href{https://osf.io/xrb76/?view_only=a8439d2b7b8548d19b65dd4e56478aab}{Supplementary Materials}) for training our tweet-level classifier. We fine-tune the pre-trained Chinese RoBERTa model~\cite{cui-etal-2020-revisiting} provided by Huggingface to perform multi-class classification. A dense layer is added on top of RoBERTa while fine-tuning. We achieve an overall F1 score (imbalanced classes) of 0.77 over 5-fold cross-validation. The performance across classes is reported in Table~\ref{tab3}.
\input{tables/classification-performance}

\subsubsection{Identifying influential/core accounts.}
We identify the \pro and \anti accounts that are most influential in driving the Chinese narratives around RU on Twitter, which we refer to as the {\it core} accounts. We manually identify 11 PRC (People's Republic of China) Official accounts in our dataset. We calculate an account's {\it influence} by the harmonic mean of its in-degree centrality and betweenness centrality in the retweet network and manually label 150 most central accounts
% as \pro accounts, \anti accounts, or \unsure 
by looking at 20 of their randomly selected tweets. This yields 87 \anti core accounts and 10 \pro core accounts which suggests that the \anti actors are more central in our retweet network.

To expand our \pro core we use a stratified sampling approach. We leverage the RoBERTa probabilities together with centrality measures to identify the \pro core accounts. Specifically, we extract accounts with indegree centrality greater than 75\% or closeness centrality greater than 70\% and manually label 150 accounts with the highest upper quartile of pro probabilities for 20 topical tweets. This results in 87 \anti core and 106 \pro core (including PRC Official) accounts. 

\subsubsection{Account-level classification.} To identify the political leanings of the narrative creators, we classify accounts as \pro or \anti leaning. Given the prevalent observations of users' homophilous retweeting behaviors, we consider the retweeting tendency as a proxy for political leaning, as calculated below: 
\begin{equation}
    leaning = \frac{\# retweet \hspace{2mm} \pro \hspace{2mm} core}{\# retweet \hspace{2mm} \pro \hspace{2mm} core + \# retweet \hspace{2mm} \anti \hspace{2mm} core}
\end{equation}

Those accounts with a leaning greater than 0.5 are considered \pro leaning, as they retweet the \pro core accounts more often than the \anti core accounts, while those with a leaning less than 0.5 are considered \anti leaning. To account for temporal variation in our analysis, we divide our dataset into two time periods for account-level classification: March 2022 (T1) and April 2022 (T2). Thus, accounts can switch leaning from T1 to T2. We validated our leaning proxy assumption by manually labeling 25 \pro leaning and 25 \anti leaning samples from each T1 and T2. We train two BiLSTM classifiers in T1 and T2 using the leaning labels as ground truth to perform binary classification in T1 and T2 (details in \href{https://osf.io/xrb76/?view_only=a8439d2b7b8548d19b65dd4e56478aab}{Supplementary Materials}). Table~\ref{tab6} shows our classifier's T1 and T2 class proportions.
\input{tables/class-proportion}

% As shown in Fig.~\ref{fig0}, 
The majority (53\%) of new accounts appearing in T2 are \anti leaning. Most of the \anti and \pro accounts in T1 remain \anti and \pro in T2 with around 20\% of accounts becoming \unsure for both classes. We find that around 8\% \pro accounts shift their leaning to \anti and 6\% \anti accounts shift their leaning to \pro from T1 to T2 using this approach.

\subsubsection{Identifying narratives.} We extract all the trigrams (having frequency more than two) from our tweets dataset. We create weekly $n$-gram ($n=3$) co-occurrence networks based on $n$-grams co-occurring in a tweet. We further create $n$-gram co-occurrence networks for \pro account, \pro core account, \anti account, and \anti core account tweets (including retweets) for each week. Using the Louvain community detection~\cite{blondel2008fast} method, we construct $n$-gram clusters from our $n$-gram co-occurrence networks and manually identify 38 most popular subtopics related to RU from the $n$-gram clusters. These subtopics captured the diverse narratives surrounding RU on Chinese Twitter.

\section{Results}

\subsection{RQ1: Chinese Political Narratives around RU}
Tweets with subtopic-related $n$-grams are mapped to the 38 subtopics. This yields 5209 (about 11\%) subtopical tweets. Fig.~\ref{fig1} shows subtopics by account leaning. The most common subtopic is ``Ukraine (U) rescue and fund new China (C)"--funding the New China Federation to rescue Ukraine. This topic is driven by \anti accounts. The \anti accounts tend to focus on topics that criticize the Chinese Communist Party (e.g., ``GreatTranslationMovement," ``Take down the CCP"), Russia and Putin, Ukraine, and China's relationships with the United States and Russia, respectively. The \pro account narratives praised China's efforts (e.g., `Chinese (C) in Ukraine (U) rescued', `China (C) will play a constructive role'), criticized the US (e.g., `West lecturing', `gun violence in the US'), Sino-US relationship, and Russia-China partnership. Both the \pro and \anti accounts contribute to the narrative building for certain subtopics. We find polarized opinions on these subtopic-related narratives. For example, the \anti accounts support US arms to Ukraine (U), while the \pro accounts call it US imperialism. 

\input{images/fig1}

Fig.~\ref{fig4}(a) shows the \pro and \anti account activity (\#tweets + \#retweets) during the 8 weeks. Weeks 4–8 show increased \anti account activity. From week 5 to week 8, \pro account activity drops. The \anti core activity however remains consistent over the weeks. We examine subtopical tweet account activity in Fig.~\ref{fig4}(b) to see if a similar trend exists. For RU-related subtopics, \pro accounts were more active in the first four weeks and less active in the later weeks, similar to the overall activity trend. Subtopic-related \anti activity increases from week 5 to 8 with a sudden increase in week 8.
\input{images/fig2}

We examine narrative evolution using the top 12 subtopics. Other interesting subtopics include ``US arms to U," ``R-C strategic partnership," ``UN Human Rights (HR) council," and ``Bucha Massacre." Fig.~\ref{fig9} shows the 16 subtopics' distribution over time. Weeks 5 and 8 saw increased activity due to the subtopic ``U rescue and fund new C." We see a wider range of subtopics in the first 4 weeks, then fewer subtopics in the later weeks.
\input{images/fig3}

\subsection{RQ2: Contribution of Camps to RU-related Narrative Evolution}
To investigate the \pro and \anti camp's contributions to narrative evolution, we look at the weekly attention received by the subtopics across these camps. Fig.~\ref{Fig:2}(a) shows the 16 subtopics' weekly frequencies in colors. Many subtopics (e.g., ``U rescue and fund new C," ``Take down CCP," ``GreatTranslationMovement") appeared most in week 8. The subtopic ``biolab in U" peaked during week 2 and disappeared after week 5, suggesting that the \pro camp moved on from the Ukraine biolabs conspiracy.

We examine \pro and \anti camps's weekly contributions to these subtopics. Each subtopic's weekly leaning score is calculated by subtracting its \pro activity from its \anti activity and dividing by their sum (\#total activity). A score near 1 indicates that \anti accounts dominated the subtopic, while a score near -1 indicates \pro accounts dominated. Only subtopics with more than three tweets in the week are considered. Fig.~\ref{Fig:2}(b) shows the weekly subtopic leanings.
\input{images/fig4}

Throughout our study, the \anti camp focused on subtopics like ``U rescue and fund new C," ``Take down CCP," ``GreatTranslationMovement," and ``dark Deep State (DS) govt." The \pro camp focused on subtopics ``biolab in U," ``Sino-US relations," and ``Putin warned Finland." The ``Sino-US relations" subtopic is polarized, with both sides contributing but the \pro side promoting it. ``R army crimes," ``Bucha massacre," and ``Jewish Holocaust massacre" were introduced by \anti camp while the \pro camp introduced the subtopic ``US arms to U". Week 2's ``biolab in U" subtopic was equally attended by \pro and \anti camps.

The \anti camp hijacked \pro camp subtopics like ``WW3", ``R-C strategic partnership," and ``US arms to U" in later weeks. When discussing these subtopics, the \anti camp often flipped the \pro camp's narratives. For the subtopic ``WW3," the \pro camp claimed that US-funded biolabs in Ukraine were making bioweapons for World War \rom{3}: {\it ``U.S.-funded Ukrainian biological weapons laboratory destroyed by Russia, the highest level of silent needle war against the people..."}. The \anti camp narrative claimed that CCP started the WW3 by ``releasing" the COVID-19 virus and {\it ``Russia’s invasion of Ukraine is just an extension of the virus war to the hot war…"}. This suggests that \anti camp actively counters the narratives of \pro camp, whereas no such observations were found from the \pro camp. Consistent with our previous findings, \pro camp activity decreased in weeks 5-8 while \anti camp activity increased.

To understand the shift of focus of the \pro camp, we look at how the 11 PRC Official accounts retweeted and were retweeted during this time period, shown in Fig.~\ref{Fig:2}(c). During the first month, PRC accounts were retweeted more on the subtopical tweets (``tweeted\_sub") as compared to the second month. The overall trend of PRC Official accounts being retweeted (``tweeted\_oth") remained consistent. Interestingly, we find that the PRC Official accounts started retweeting (``retweet\_oth") from week 5, most of which were related to Taiwan. This suggests that the \pro core accounts started actively pursuing a different topic which may be the reason the \pro camp shifted focus from RU-related narratives.

\section{Discussion and Future Work}
We studied the evolution of RU-related narratives in the Chinese Twitter space and how the anti-state and pro-state actors contributed to the narrative evolution. We found that propagandist narratives arise in both the pro-state and anti-state camps. Our study reveals that the Chinese Twitter space is dominated by anti-state voices. Both these findings are consistent with previous studies~\cite{bolsover2019chinese}. 

The results showed that a few influential accounts are responsible for introducing the propagandist narratives in both pro-state and anti-state camps which are then spread (through retweets) by the rest of the camp members. This is a core aspect of participatory propaganda where an information campaign is seeded by strategic actors and spread to the mainstream public by co-opting community members~\cite{lewandowsky202220}. It was seen that the pro-state camp members shifted their focus from RU-related narratives when the pro-state core accounts actively started pursuing other topics (e.g., Taiwan). Nevertheless, understanding the exact strategies behind how a disinformation cascade unfolds requires more analysis and we leave that as a part of the future work.

We found that the pro-state influential accounts mostly focus on criticizing the US and anti-state influential accounts on criticizing the CCP. There were instances of polarized narratives by the pro-state and anti-state camps on the same subtopics to achieve their respective propaganda goals. Moreover, our analysis revealed that certain subtopics initiated by the pro-state camp were more likely to be appropriated by the anti-state camp in subsequent weeks, whereas the converse was less likely, suggesting that the pro-state camp may engage less actively in countering its critics, which resonates with previous studies~\cite{king_pan_roberts_2017}. This may suggest that the two groups have distinct offensive-defensive strategies, with one group focusing more on initiating novel propagandist narratives than the other. Future research may investigate the plausibility of the hypothesis in a different national or international context.

One of the limitations of our study is the short-term observation of the community activity---especially the lack of observation before the Russian invasion of Ukraine. Future work can look into the longer-term evolution of key communities in the pro-state and anti-state networks and community contribution to narrative building. 

\section*{Supplementary Materials}
% \vspace{1em}
\paragraph{\bf Ukraine biolab-related keywords.} We use the following keyword list to extract tweets related to the Ukraine biolabs conspiracy: 
\chinese{``烏 生化實驗室",
``烏 生物實驗室",
``烏 生化武器",
``烏 生物武器",
``烏 生化实验室",
``烏 生物实验室",
``烏 生化武器"}, and
\chinese{``烏 生物武器"}
 (``U Biochemical Laboratory", ``U Biological Laboratory", ``U Biochemical Weapon", ``U Biological Weapon", ``U Biochemical Laboratory", ``U Biological Laboratory", ``U Biochemical Weapon", ``U Bioweapon"), where space between words denotes AND.
 
\paragraph{\bf Identifying topics.} During the manual labeling process, we identify 8 broad (overlapping) tweet topics namely `Russia-Ukraine War', `Taiwan', `COVID Origins', `Tibet', `US Topics', `Hong Kong', `Xinjiang', and `Tiananmen' and a set of keywords related to each topic. The set of keywords is refined during the manual labeling process. We refer to the tweets containing these topic-related keywords as topical tweets.
\paragraph{\bf Account Classification details.} We select accounts that have retweeted the core minimum of three times and have over three Chinese language tweets (including retweets) in our dataset. This yields 1381 accounts in T1 (29\% \pro leaning) and 2739 accounts in T2 (15\% \pro leaning). We train (80-20 split) XGBoost, SVM, Logistic Regression, Random Forest, Gradient Boost, and BiLSTM classifiers in T1 and T2 using stratified bootstrapped sampling with 40\% \pro training samples and prediction probabilities are averaged over 4 iterations. We get the RoBERTa \pro and \anti probabilities and the tweet embedding vectors (dimension=256) of 20 randomly selected topical tweets for each account. The BiLSTM model is trained on fasttext word embeddings after concatenating the 20 tweets (sorted according to the maximum of their \pro and \anti probabilities). For the shallow models, our feature set includes, the mean embedding vector, mean \pro probability, upper quartile of \pro probability, mean \anti probability, and upper quartile of \anti probability. The account-level classifier performances in T1 and T2 are reported in Table~\ref{tab4}.
\input{tables/account-level-performance}

We choose the BiLSTM models for classification in both T1 and T2 to maintain consistency. We perform an error analysis of BiLSTM predictions of the validation set to determine the ideal probability cutoffs for \pro and \anti classification. The accounts having probability less than 0.1 are considered \anti leaning, above 0.9 are considered \pro leaning, and the rest are considered as \unsure accounts. We use our classifiers to classify the leaning of the remaining accounts (having more than three Chinese-language tweets) in T1 and T2.

\section*{Acknowledgement}
The authors gratefully acknowledge the support from AFOSR, ONR, Minerva, and Pitt Cyber Institute's PCAG. This research was supported in part by the resources provided by the University of Pittsburgh Center for Research Computing, RRID:SCR\_022735, through support by NIH \#S10OD028483. Any opinions, findings, and conclusions or recommendations expressed in this material do not necessarily reflect the views of the funding sources.

% \vspace{-2mm}

{
\renewcommand{\clearpage}{}
\renewcommand{\bibname}{\Large\bfseries References}
\renewcommand\bibpreamble{\vspace{-3.0\baselineskip}} 
\setlength{\bibsep}{0pt} 
{\small \bibliography{samplepaper}}
}

\end{document}

%% file: tables/classification-performance.tex
\begin{table}
\vspace{-1em}
\centering
\setlength{\tabcolsep}{4pt}
\caption{Performance summary of tweet-level classifier across classes.}\label{tab3}\vspace{-1em}
\begin{tabular}{lccc}
\toprule
\textbf{class} & \textbf{precision} & \textbf{recall} & \textbf{F1}\\
\midrule
\anti & 0.80 & 0.93 & 0.86 \\
\pro & 0.69 & 0.69 & 0.69 \\
\unsure & 0.81 & 0.51 & 0.63 \\
\bottomrule
\end{tabular}\vspace{-1.5em}
\end{table}

%% file: tables/class-proportion.tex
\begin{table}[h]\vspace{-1.5em}
\centering
\setlength{\tabcolsep}{5pt}
\caption{Proportion of classes using account-level classifier in T1 and T2}\label{tab6}\vspace{-1em}
\begin{tabular}{lcccc}
\toprule
\textbf{setting} & \textbf{\#\pro} & \textbf{\#\anti} & \textbf{\#\unsure} & \textbf{\#samples}\\
\midrule
T1 train, T1 test & 488 (12\%) & 1743 (43\%) & 1781 (45\%) & 4012\\
T2 train, T2 test & 1139 (21\%) & 1751 (33\%) &  2411 (46\%) & 5301\\

\bottomrule
\end{tabular}\vspace{-1.5em}
\end{table}

%% file: images/fig1.tex
\begin{figure*}[h]
\vspace{-1em}
\includegraphics[width=\textwidth]{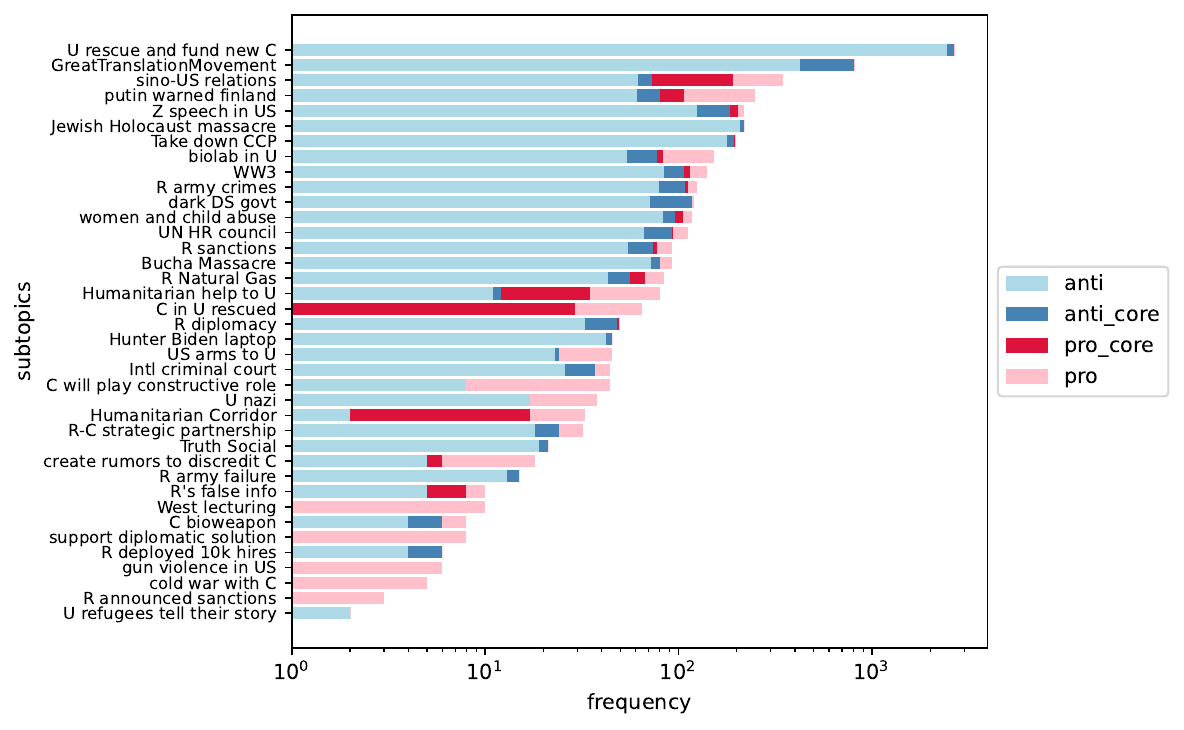}
\vspace{-2.5em}
\caption{Frequency of subtopics across account leaning} \label{fig1}
\vspace{1.2em}
\end{figure*}

%% file: images/fig2.tex
\begin{figure*}[h]
\vspace{-1em}
\centering
\setlength{\tabcolsep}{0pt}
\begin{tabular}{cc}
(a) & (b)\\
\includegraphics[width=.5\textwidth]{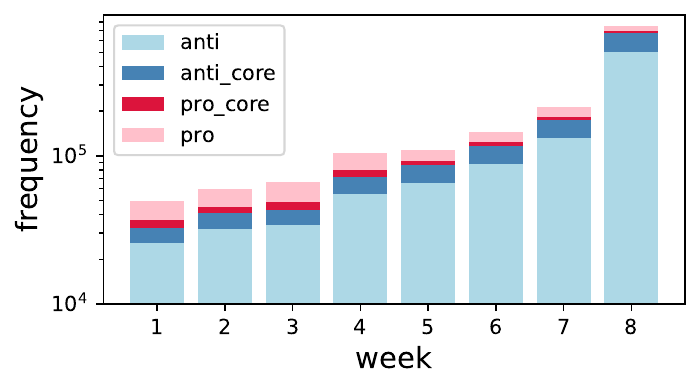}
&
\includegraphics[width=.5\textwidth]{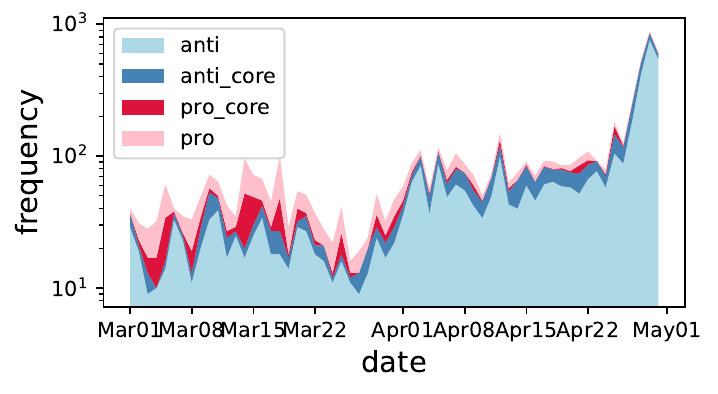}\\
\end{tabular}\vspace{-1.5em}
\caption{(a) Overall activities (b) Activities with respect to the subtopical tweets} 
\label{fig4}
\end{figure*}

%% file: images/fig3.tex
\begin{figure*}[h]\vspace{-1.5em}
\centering
\includegraphics[width=\textwidth]{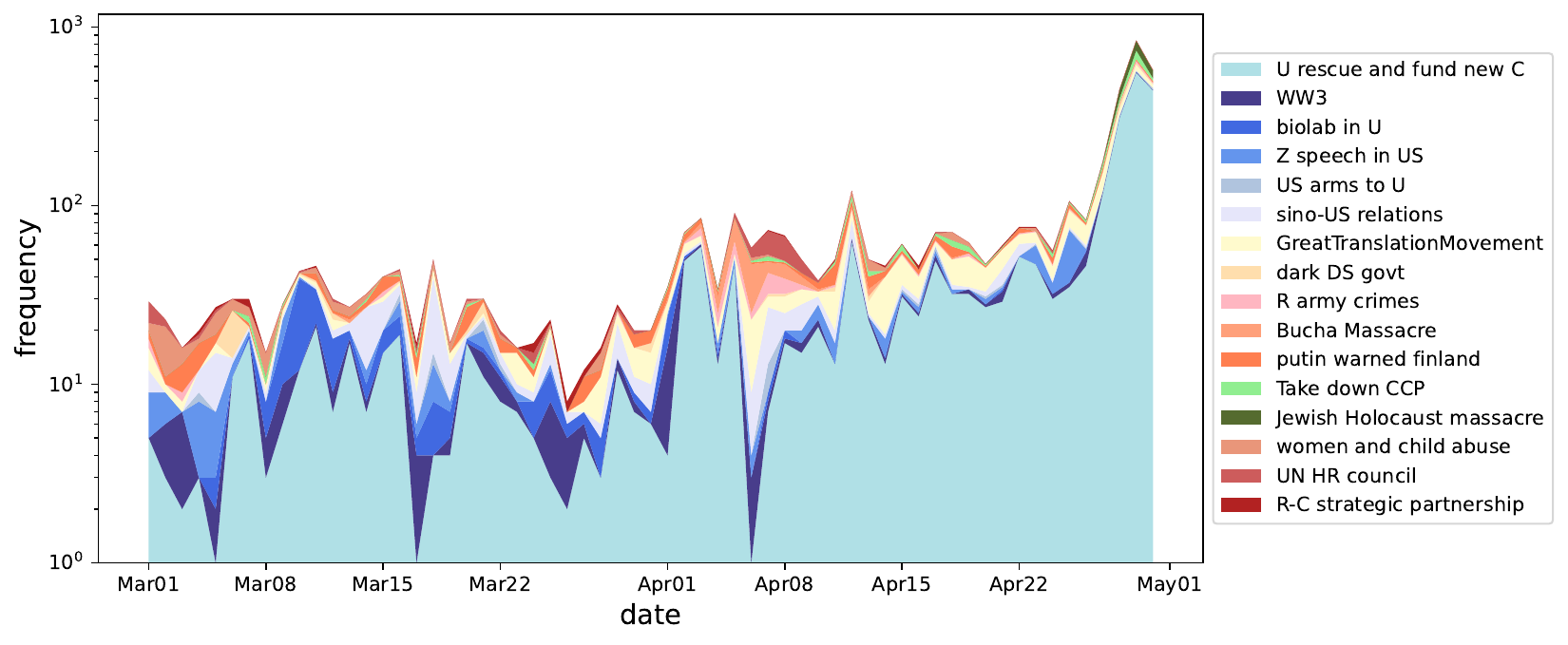}\vspace{-1.8em}
\caption{Distribution of 16 selected subtopics}
\label{fig9}
\end{figure*}

%% file: images/fig4.tex
\begin{figure*}[h]
\vspace{-1.5em}
\centering
\setlength{\tabcolsep}{0pt}
\begin{tabular}{cc}
(a) & (b)\\
\includegraphics[width=.5\textwidth]{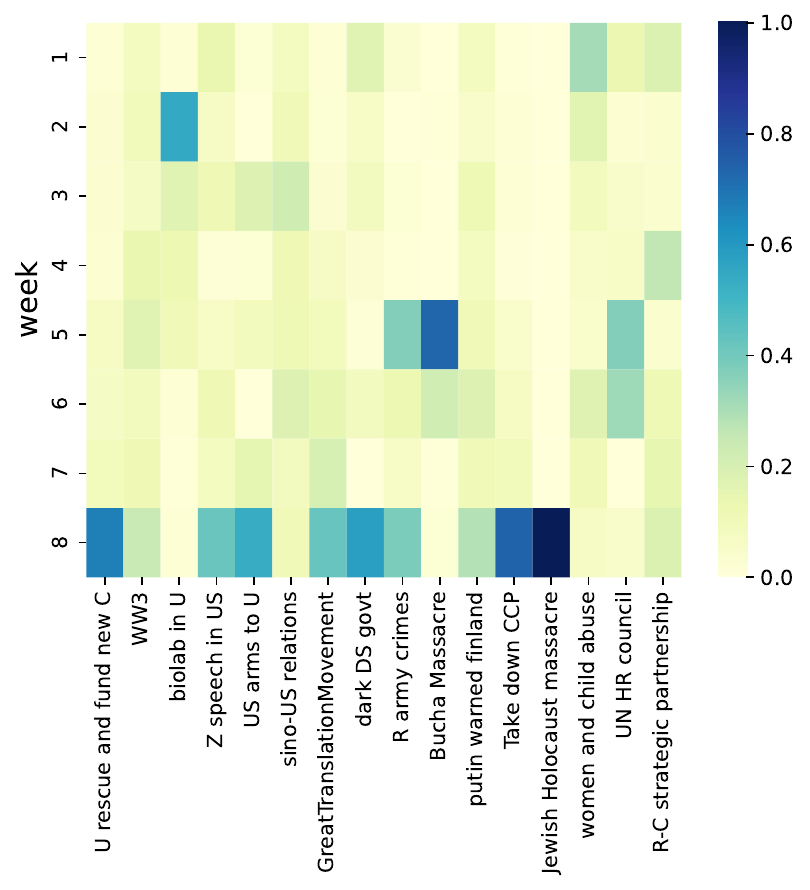}
&
\includegraphics[width=.5\textwidth]{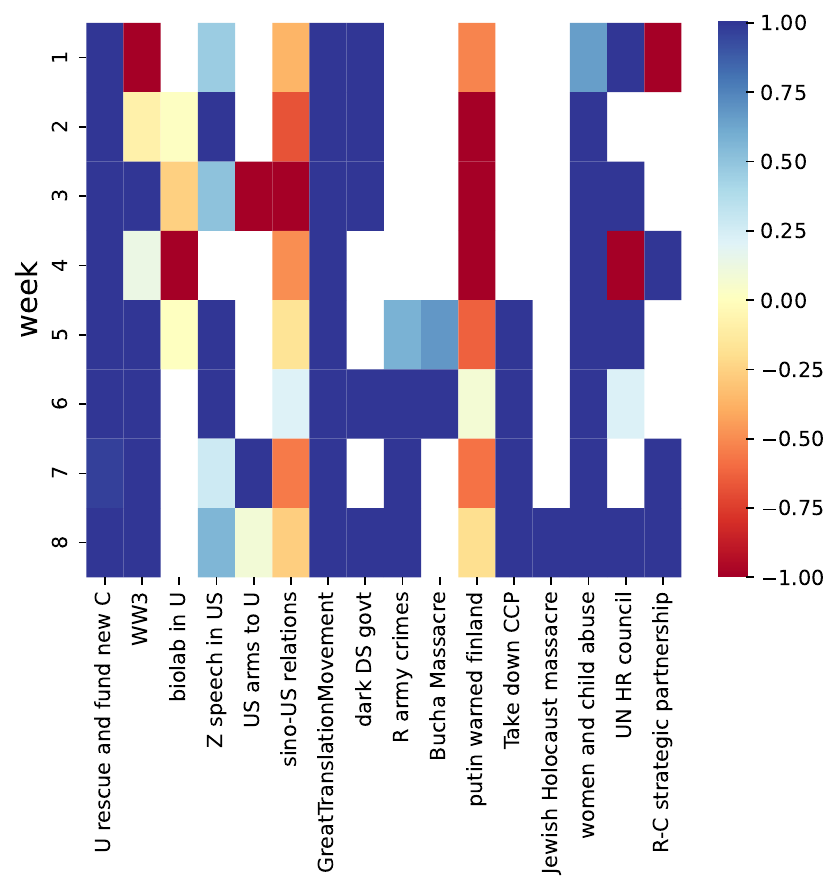}\\
\multicolumn{2}{c}{(c)}\\
\multicolumn{2}{c}{\includegraphics[width=.7\textwidth]{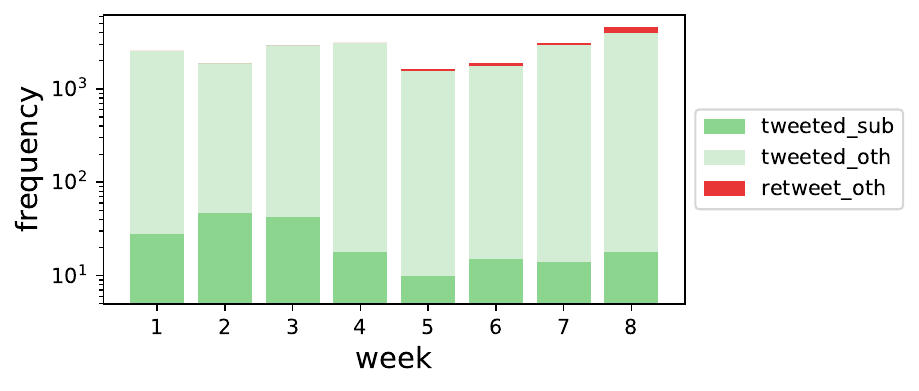}}\\
\end{tabular}\vspace{-1.5em}
\caption{Weekly changes of (a) distribution of subtopics, (b) subtopic leanings, and (c) PRC Official activity.}
\label{Fig:2}
\vspace{-0.7em}
\end{figure*}

%% file: tables/account-level-performance.tex
\begin{table}\vspace{-1em}
\centering
\setlength{\tabcolsep}{3.5pt}
\caption{Performance summary of account-level classifier in T1 and T2.}\label{tab4}
\begin{tabular}{lcccccccc}
% {|l|l|l|l|}
\toprule
 & \multicolumn{4}{c}{\textbf{T1}} & \multicolumn{4}{c}{\textbf{T2}} \\
\cmidrule(rl){2-5} \cmidrule(l){6-9}
\textbf{model} & \textbf{acc\%} & \textbf{precision} & \textbf{recall} & \textbf{F1} & \textbf{acc\%} & \textbf{precision} & \textbf{recall} & \textbf{F1}\\
\cmidrule(r){1-1} \cmidrule(rl){2-5} \cmidrule(l){6-9}
LR & 88 & 0.89 & 0.88 & 0.88 & 90 & 0.92 & 0.90 & 0.91 \\
SVM & \textbf{89} & \textbf{0.90} & \textbf{0.89} & \textbf{0.90} & 90 & 0.91 & 0.90 & 0.90 \\
Random Forest & 87 & 0.87 & 0.87 & 0.87 & 90 & 0.91 & 0.90 & 0.91 \\
Gradient Boost & 87 & 0.88 & 0.87 & 0.87 & 90 & 0.91 & 0.90 & 0.90 \\
XGBoost & 85 & 0.85 & 0.85 & 0.85 & 91 & 0.92 & 0.91 & 0.91 \\
BiLSTM & 86 & 0.86 & 0.86 & 0.86 & \textbf{94} & \textbf{0.93} & \textbf{0.94} & \textbf{0.93} \\

\bottomrule
\end{tabular}\vspace{-1em}
\end{table}